\documentclass[
  journal=pasa,
  manuscript=research-paper, 
  year=2023,
  volume=40,
]{cup-journal}

\usepackage{microtype,siunitx,booktabs}
\usepackage{multirow}
\usepackage{xcolor}
\usepackage{ulem}
\usepackage{amsmath}
\usepackage{pdflscape}
\allowdisplaybreaks

\newcommand{\best}[1]{\textit{\textbf{#1}}}
\newcommand{\eat}[1]{}
\newcommand{\vit}{\textsc{v}\textbf{\scriptsize{i}}\textsc{t}}

\newcommand{\rgzcomment}[1]{{\color{cyan}{\bf RGZ-Team:~}{#1}}}
\newcommand{\daweicomment}[1]{{\color{blue}#1}}
\newcommand{\daweidel}[1]{\daweicomment{\sout{#1}}}
\newcommand{\vinaycomment}[1]{{\color{red}#1}}
\newcommand{\vinaydel}[1]{\vinaycomment{\sout{#1}}}

\renewcommand{\rgzcomment}[1]{}
\renewcommand{\daweicomment}[1]{#1}
\renewcommand{\daweidel}[1]{}
\renewcommand{\vinaycomment}[1]{#1}
\renewcommand{\vinaydel}[1]{}

\title{
Radio Galaxy Zoo: 
Tagging 
Radio Subjects using Text
}

\author{Dawei Chen}
\affiliation{joint first authors}
\alsoaffiliation{School of Computing, Australian National University (ANU)}

\author{Vinay Kerai}
\affiliation{joint first authors}
\alsoaffiliation{University of Western Australia (UWA)}

\author{Matthew J. Alger}
\affiliation{Google Australia}

\author{O.\ Ivy Wong}
\affiliation{CSIRO Space \& Astronomy, PO Box 1130, Bentley, WA 6102, Australia}
\alsoaffiliation{ICRAR-M468, University of Western Australia, Crawley, WA 6009, Australia}
\alsoaffiliation{ ARC Centre of Excellence for All-Sky Astrophysics in 3 Dimensions (ASTRO 3D), Australia}

\email[O.\ Ivy Wong]{ivy.wong@csiro.au}

\author{Cheng Soon Ong}
\affiliation{Data61, CSIRO}
\alsoaffiliation{School of Computing, Australian National University (ANU)}

\doi{10.1017/pasa.2020.32}

\received {dd Mmm YYYY}
\revised  {dd Mmm YYYY}
\accepted {12 Oct 2023}
\published{dd Mmm YYYY}


\begin{document}

\begin{abstract}
RadioTalk is a communication platform that enabled members of the Radio Galaxy Zoo (RGZ) citizen science project to engage in discussion threads and provide further descriptions of the radio subjects they were observing in the form of tags and comments. It contains a wealth of auxiliary information which is useful for the morphology identification of complex and extended radio sources. In this paper, we present this new dataset, and for the first time in radio astronomy, we combine text and images to automatically classify radio galaxies using a multi-modal learning approach. We found incorporating text features improved classification performance which demonstrates that text annotations are rare but valuable sources of information for classifying astronomical sources,
and suggests the importance of exploiting multi-modal information in future citizen science projects. We also discovered over 10,000 new radio sources beyond the RGZ-DR1 catalogue in this dataset.
\end{abstract}


\section{INTRODUCTION}
\label{sec:intro}

Widefield radio astronomy surveys using the Square Kilometre Array (SKA) pathfinder instruments \citep[e.g.\ ][]{hotan21,hurley-walker17,jarvis16} are heralding an era of transformational change in methods for data processing and analysis thanks to the vast data rates and volumes resulting from these new facilities. For example, the Australian Square Kilometre Array Pathfinder \citep[ASKAP; ][]{hotan21} has already mapped approximately 2.1 million radio sources in the observatory-led Rapid ASKAP Continuum Survey \citep[RACS;][]{hale2021racs}, and the upcoming
Evolutionary Map of the Universe (EMU) survey is expected to map approximately 40 million \citep{norris21}. How to analyse such a significant number of radio sources has now become an active area of study in itself. For example, the SKA project has issued source-finding data challenges as a part of their science preparatory activities \citep{bonaldi18,bonaldi21}. Other problems include the aggregation of discontinued emission components, and classification of radio galaxies into physically meaningful classes~\citep[Sec 2.4.2]{alger21phd}.

Radio galaxies come in various shapes with various different physical properties. A well known example of radio galaxy morphology types is Fanaroff-Riley Type I and II galaxies~\citep{fanaroff1974}, whereby Type I represents jets which are brighter in the core regions and
Type II represents jets that are brighter toward the ends (called ``hot spots'').

Identifying radio galaxy morphologies en masse is very difficult. This is because of both the large diversity of radio galaxy morphologies and the variation in appearance due to different observational constraints. Radio galaxies are challenging objects: The extents of their complex structures are not necessarily bound to the stellar components of their host galaxies and can range from parsecs to megaparsecs in scale. Due to this difficulty, radio galaxy morphologies are traditionally identified through visual examination by expert astronomers
\citep[e.g. for the G4Jy Sample; ][]{white2020a,white2020b}.
Automated radio source classifiers based on machine learning methods have risen in prominence in recent years thanks to the increase in data and computing power availability \citep[e.g.\ ][]{polsterer15,aniyan17,alger18,wu19}.
Note that the morphology identification problem is also often called `classification' of the galaxies. We avoid this phrasing throughout this paper to avoid confusion with the machine learning concept of classification.

Radio Galaxy Zoo (RGZ) is an online citizen science project that asked volunteers to 1) associate disconnected radio source components, and 2) match these to their host galaxies  \citep{banfieldRadioGalaxyZoo2015,wong23}. The RGZ website would show volunteers coordinate-matched radio and infrared (IR) images of extended radio sources. See~\ref{appendix:rgzinterface} for an illustration of the online interface presented to volunteers \citep{banfieldRadioGalaxyZoo2015}. RGZ Data Release 1  \citep[henceforth DR1; ][]{wong23} catalogues the associations and host galaxies for 98,559 sources from the Faint Images of the Radio Sky survey \citep[FIRST; ][]{white97}. 
In addition to these core tasks of associating and matching, citizen scientists were also able to tag and comment on the radio sources they were labelling.
%

\begin{landscape}
\begin{figure}[p]  
\centering
\includegraphics[width=\textwidth]{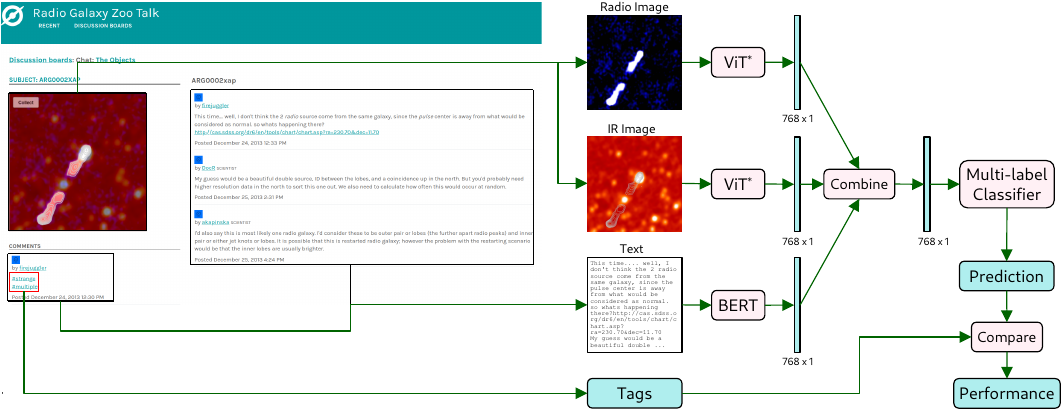}
\caption{
\rgzcomment{the general guideline for text in figures is that the size should be at least comparable to the size of main text.}
\rgzcomment{a set of example tags on the cyan tag box would go a long way to help differentiate them from the text input to BERT.}
Overview of the machine learning workflow for classifying radio sources in this paper. We apply the same pre-trained \textbf{Vi}sion \textbf{T}ransformer~\citep{dosovitskiy2020image}, denoted as ViT$^*$ in the workflow, to both the radio and infrared (IR) images to produce their numerical presentations, and the pre-trained BERT language model~\citep{devlinBERTPretrainingDeep2019} is used to create a numerical representation of the corresponding text discussions on RadioTalk forum after removing the tags. These representations are combined before passing it to a multi-label classifier which predicts tags likely applicable to the radio source. The classification performance is obtained by comparing the ground truth tags in the RadioTalk forum text and the tags predicted by the classifier. The screenshot in this figure shows images and discussions text for subject ARG002XAP.
} 

\label{fig:astrotag-workflow}
\end{figure}
\end{landscape}

\noindent
Both citizen scientists and the professional science team discussed these sources through the online ``RadioTalk'' forum. This had profound benefits, such as the discovery of a previously unknown bent giant radio galaxy \citep{banfield16}. 
It has been found that the RadioTalk forum and the further comparisons to other ancillary observations were necessary to 
identify and classify these very large sources~\citep{banfieldRadioGalaxyZoo2015,banfield16}.

The RGZ Data Release 1 (RGZ-DR1) catalogue and its predecessors are derived from the core labelling
task \citep{wong23}, and has been used in the development of a variety of machine learning radio source classifiers \citep{lukic18,alger18,wu19,galvin19,ralph19,tang22,slijepcevic2023radio}. Building upon
these investigations, we present in this paper a radio source classifier that is jointly trained on both images from RGZ-DR1 and text from the RadioTalk forum.  

The RadioTalk forum \vinaydel{enabled}\vinaycomment{allowed} volunteers to \vinaycomment{optionally contribute further}\vinaydel{describe in detail their} observations for specific radio subjects \vinaycomment{by assigning tags, writing comments and interacting via various discussion boards}\vinaydel{through contributions of tags, comments and topics for discussion on the various discussion boards}. RadioTalk, therefore, contains a wealth of auxiliary information currently not present in the RGZ catalogue. Since there has been no previous analysis of the forum or tag data, this posed a unique opportunity for our work to utilise techniques in machine learning to extract new insights, improve future information retrieval and maximise the science output from the RGZ project. Furthermore, the use of the text data from the RadioTalk forum is likely to benefit the $\approx30$~percent of the initial input sources for which the resulting classifications were too uncertain for inclusion into the RGZ-DR1 catalogue.  

Tags have also been discussed in recent studies as the way forward for future citizen science projects that are aimed at classifying radio galaxy morphologies \citep{rudnick2021}. From an annotated sample of radio sources, \citet{bowles22} and \citet{bowles23} explored the derivation of semantic classes as a multi-modal problem for the purpose of providing more accurate descriptions of radio source morphologies.
%
Within the field of astronomy, natural language processing (NLP) models have also been used to explore the text from astronomy publications. For example, \citet{ciucua2023galactic} investigated the potential of pre-trained large language models for comparing and summarising astronomical studies, or proposing new research ideas.

For the first time in radio astronomy, we combine images and text to automatically identify radio galaxy morphologies. Such an approach builds upon the merits of using tags and text  \citep{rudnick2021, bowles22, bowles23}, while retaining the advantage of the spatial information from the image datasets.  Specifically, our work focuses on the tagging functionality of the forum used by the volunteers to assign concise descriptions to the subjects they were observing. These tags are particularly valuable for extended radio sources which are of high interest to radio astronomers. A suitable classification scheme for extended radio sources is still debated amongst astronomers as quantitative classification schemes alone are insufficient in capturing the diverse set of jet morphologies. Subjects that share similar tags may enable better categorisation of sources that share physical phenomena. 

In this paper, we introduce the RadioTalk dataset. This is a dataset of over 10,000 manually labelled images of complex radio objects, as well as text features that represent these objects. This dataset is useful for both astronomers who are interested in radio objects, and machine learning researchers who are interested in real combined image/text datasets on which to demonstrate novel machine learning methods. We use this dataset to demonstrate the utility of a combination of text and image features for automated classification of complex radio objects. We train a combined image and text model to predict ``hashtags'' for complex radio sources, with our workflow presented in Figure~\ref{fig:astrotag-workflow}.
We show that machine learning \eat{approaches }on forum text provides a useful approach for galaxy classification,
and complements image based approaches.

Section 2 describes the datasets used in this paper, while Section 3 describes the machine learning workflow.  We present our results and discussion in Sections 4 and 5, respectively.  Section 6 concludes with a summary of our work.

\section{RADIO GALAXY ZOO DATASETS}
\label{sec:dataset}
The RGZ-DR1 catalogue connects associated radio components to their originating host galaxies for subject classifications with a minimum of 0.65 (\daweicomment{user-weighted consensus level}), which has a statistical reliability of at least 75\% \citep{wong23}. 
While this catalogue is not central to the topic of this paper, it was useful as an independent avenue for validating the image and text data that is described in this section.

\subsection{RGZ radio and infrared images}
\label{sec:image_data}

Coordinate-matched radio and infrared images are crucial for the core task of radio source classification due to the expected physical spatial symmetry of synchrotron emission from radio sources and their host galaxies, even if the radio source morphologies and extents are loosely constrained, and dependent on instrumental sensitivity, resolution and other imaging systematics.  The infrared images traces the stellar population of the host galaxies from which the radio emission could originate.   As the radio emission can be spatially offset from its originating host galaxy, visual inspection from citizen scientists help increase the efficiency of such classification tasks.

RGZ matches radio images from the  Faint Images of the Radio Sky at Twenty Centimeters \citep[FIRST; ][]{white97} and the Australia Telescope Large Area Survey Data Release 3 \citep[ATLAS; ][]{franzen15} surveys to infrared images from the Wide-Field Infrared Survey Explorer \citep[WISE; ][]{wright10} and the Spitzer Wide-Area Infrared Extragalactic Survey \citep[SWIRE; ][]{lonsdale03} surveys, respectively.  As shown in Figure~\ref{fig:astrotag-workflow}, the WISE 3.4~$\mu$m ($W1$) infrared image is presented as a heatmap that is overlaid with contours of the radio emission.  These radio contours begin at the 4$\sigma$ level with increments of factors of $\sqrt{3}$.  We note that within the RGZ project's interface (as illustrated in~\ref{appendix:rgzinterface}), these radio contours can be faded into a blue-scaled image but for the purpose of this paper, we use the single heatmap IR images that has radio contours overlaid.

In this paper, we use the 10,643 radio and heatmap images from the pre-release DR1 catalogue that was also used by \citet{wu19}, and extract features of both the radio contours that are overlaid on the infrared heatmap images with pre-trained vision transformer~\citep{dosovitskiy2020image}. In particular, the features of a radio image is a 768-dimensional embedding, and similarly for the corresponding heatmap images. These embeddings, together with those of the corresponding text discussions, are the inputs of a feedforward neural network (i.e. MLP) to classify the tags.

\subsection{Threaded text discussions from RadioTalk}
\label{sec:radiotalk}

At the conclusion of the core RGZ cross-matching task, the participants are asked if they would like to "discuss"
the subject further and tag the subject in question with hashtags.   In addition, the RadioTalk forum also includes longer-form general science discussions that are less subject-specific but could be associated with a set of subjects. The screenshot in Figure~\ref{fig:astrotag-workflow} shows an example of both the shorter discussion comments (left panel) and the longer discussions (right panel) that relate to the subject ARG002XAP. 

Due to the free-form nature of such discussions in addition to the tagging of subjects being an optional task, the associations of discussions to subjects and the tagging of subjects are likely to be highly incomplete.  Nevertheless, we found comments and tags for an additional 10,810 RGZ subjects that are not included in RGZ-DR1. 
This suggests that participants tended to discuss subjects with complex radio morphologies that do not attain the consensus level cut required by RGZ-DR1. 
We refer the reader to Figure~\ref{fig:venn_diagram} for more details and Section~\ref{subsec:complexsources} for a further discussion.

The consistency of the hashtags is further complicated by the fact that each participant could freely generate new hashtags as the use of \eat{the set of }suggested hashtags is not enforced.  This results in cases where subjects could have been tagged with a relevant hashtag that has been suggested, but in reality remains untagged.  This scenario contributes towards a positive-unlabelled dataset which has implications for the application of machine learning techniques.
%
The 20 most frequently used tags in the RadioTalk dataset can be found in Figure~\ref{fig:tag_usage}.



\subsection{Radio subject hashtags for supervised learning}
\label{sec:text}


To prepare the dataset for our machine learning classifiers we performed two main pre-processing steps. 

Firstly, we concatenated all thread text associated with a given subject.
%
Secondly, we cleaned the tags to increase coherence across the forum. 
The dataset contains over $1,000$ unique raw tags, largely containing synonymous or misspelt terms. To maximise the amount of information we can extract from the dataset we generated a mapping function from the raw tags to their processed counterpart. Our mapping function converted the raw tags to lower-case and removed special characters. To handle misspelt and abbreviated tags, we used a Levenshtein distance heuristic alongside manual revisions.
%
%
The Levenshtein distance between two tags is the minimum number of edits (i.e. insert, delete or replace a single character) required to change one tag into the other~\citep{navarro2001guided}. It is widely used to measure the difference between two text sequences, in particular for short text like the tags in our dataset.
As an explicit example, our generated mapping function maps the raw tags `Doublelobe', `double-lobe` and `doubblelobe' to the single tag `doublelobe'. This method of processing preserves semantic content while improving tag coherence across subjects and reducing the set of distinct tags.

\begin{figure}[tb]
\centering
\begin{tabular}{r}
\includegraphics[width=0.62\textwidth]{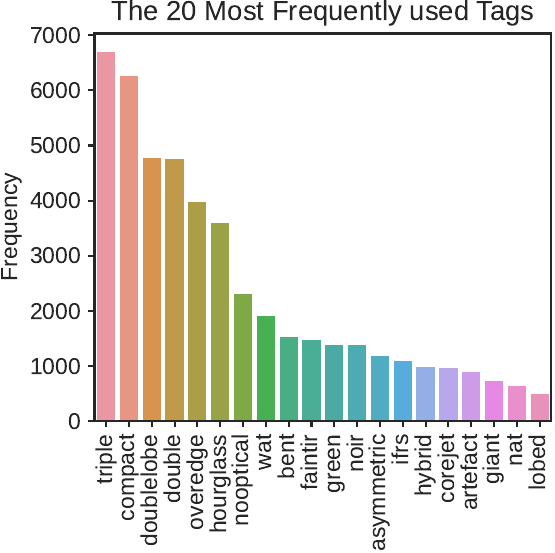} \\ 
\vspace{-2pt} \\
\includegraphics[width=0.6\textwidth]{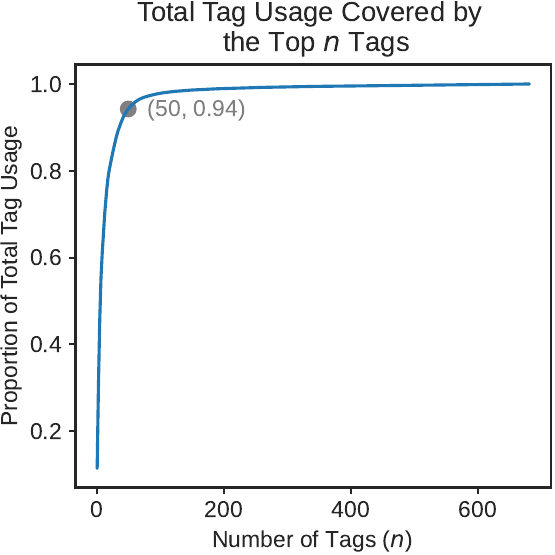}
\end{tabular}
\caption{
Summary of tag usage in the RadioTalk dataset.
\textbf{Top}: Histogram of the 20 most frequently used tags.  
\textbf{Bottom}: Proportion of the total tag usage covered by the top $n$ tags (i.e. the $n$ most frequently used tags). We can see that the top 50 tags cover 94\% of all tag usage in the RadioTalk dataset.
}
\label{fig:tag_usage}
\end{figure}

\begin{figure}[tb]
\centering
\includegraphics[width=.8\linewidth]{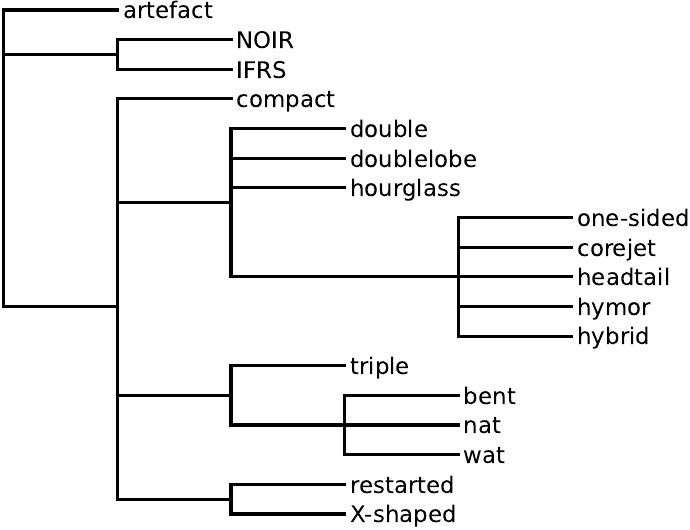}
\caption{
\
\daweicomment{Hierarchical tree of tag clustering created by an astronomer upon closer examination of the most frequently used tags in the RadioTalk dataset.}
The merger of tags in Section~\ref{sec:text} are performed according to this tree.} 

\label{fig:tagtree}
\end{figure}

After applying these steps we were left with several hundreds of tags, from which we selected the 50 most frequent tags as we found these made up 
\daweicomment{94\%} of all tag usage in the dataset (see Figure \ref{fig:tag_usage}).\rgzcomment{Given that figure 3 shows that the 50 tags make up 94\% of the total instances, I would probably just say 94 in the text.}\rgzcomment{I am not sure I agree with the statement "Additionally tags that are used frequently are likely to be most useful for astronomers." My concern around this statement echos your (very good discussion) on open classification. I think removing this line or phrasing it around something more specific would be good.}\daweidel{Additionally tags that are used frequently are likely to be most useful for astronomers.}
Upon closer examination by an astronomer (OIW), we found many tags could be describing similar morphologies, and a hierarchical tree of tag clustering (see Figure~\ref{fig:tagtree}) was created. Hence we created a smaller set of tags by merging tags with shared meaning into a single tag. 
We note that the hierarchical tree used in this paper is to group similar source structures observed from the FIRST images, for the purpose of developing an automated classifier, rather than the representation of child and parent nodes.  The merged classes may not fully represent our astrophysical understanding of the source classes.  For example, headtail sources may in fact be NATs but because the tails may not be distinguishable from the FIRST image, hence the headtail class is a more accurate description of the image.  

We created a new set of 11 tags by merging a few tags with the same meaning. The merging is performed according to the hierarchical tree of tag clustering shown in Figure~\ref{fig:tagtree}.
Table~\ref{tab:merge_tags} shows the 4 sets of tags that are merged to create 4 new tags.
In particular, if a radio subject is labelled by two or more (old) tags which will be merged into a new tag, we update the set of tags for the subject by adding the new tag and removing the old tags. For example, if a radio subject is labelled both \texttt{onesided} and \texttt{hybrid} by participants of the RadioTalk forum, according to Table~\ref{tab:merge_tags}, we would then remove both \texttt{onesided} and \texttt{hybrid} from its tags, and add \texttt{asymmetric} to its tags.
%


\daweicomment{
We argue that it is beneficial or even necessary to merge tags given the current data volume and quality, and our work benefits from having larger sample sizes per tag that results from the merger of tags. The merging scheme presented in Table~\ref{tab:merge_tags} are for technical reasons of sample sizes. It would have been nice to not have merged those tags for astrophysical reasons, or adopt alternatively approaches~\citep{bowles22,bowles23} should sufficient samples be available.
}

The set of 11 tags after merge are:
\texttt{artefact}, 
\texttt{asymmetric}, 
\texttt{bent}, 
\texttt{compact}, 
\texttt{double}, 
\texttt{hourglass}, 
\texttt{noir\_ifrs}, 
\texttt{overedge}, 
\texttt{restarted}, 
\texttt{triple}, 
\texttt{xshaped}. The \texttt{noir\_ifrs} tag is an abbreviated tag that represents the ``No Infrared / NoIR'' and the ``Infrared Faint Radio Source'' classes of radio sources whose host galaxies are not visible in the corresponding infrared or optical maps~\citep{norris2006deep}.
Descriptions of the other tags as well as
examples of radio subjects for each of the 11 tags can be found 
in~\ref{appendix:examples}.
\vspace{1em}

\begin{table}[!h]
\caption{Merging 4 sets of tags to create 4 new tags.}
\begin{tabular}{p{2.2in}l}
\toprule
Tags (before merging) & Tag (after merging) \\
\midrule
onesided, corejet, headtail, hymor, hybrid, asymmetric & asymmetric \\
NOIR, IFRS & noir\_ifrs \\
double, doublelobe & double \\
bent, nat, wat & bent \\
\bottomrule
\end{tabular}
\label{tab:merge_tags}
\end{table}

\subsection{Summary of the galaxy tagging dataset}

We shall now present an analysis of our constructed RadioTalk dataset to better understand the available data. 
\daweicomment{In total, the citizen scientists used 702,139 words within 53,065 threaded comments to discuss 34,015 radio subjects in the RadioTalk forum.}
Table \ref{tab:datasetbreakdown} shows 
the statistics of tags and comments across the threaded text discussions of radio subjects.
From this, we can deduce that the threads are very short with less than 2 comments per subject on average. In fact, most subjects contain only a single tag and a single comment. 
Such sparsity in data poses further challenge to a tag prediction system since we have little data to learn from.

\daweicomment{
After the pre-processing steps described in Section~\ref{sec:text}, we found 10,643 subjects in the RadioTalk dataset with both radio and infrared images (with radio contours) available in the RGZ-DR1 catalogue.
To create data splits for training and testing, we first sort these subjects by their unique Zooniverse ID which also sorts them chronologically. We then use the first 85\% of rows as the training data with the remainder to be used as the test set. The same splitting approach is used to create a training set and a validation set from the training data. This results in 7,688 (72\%), 1,357 (13\%) and 1,598 (15\%) radio subjects in the training set, validation set and test set, respectively, in the dataset.}
The training set is used to learn the classification models (see Section~\ref{sec:classifiers}), the validation set is for tuning the hyper-parameters of classifiers (for example, the architecture details shown in Figure~\ref{fig:clf_arch}), and the test set is reserved for evaluating the trained classifiers so that the reported classification performance is a good indication of how well the models will perform on new radio subjects.


We created a dataset archive
which is available to download from Zenodo at \texttt{\small\url{https://zenodo.org/record/7988868}}.
%
%
For each of the training, validation and test set, we provide a CSV file where each row represent a radio subject with a particular zooniverse ID ({\tt\small zid}). The columns in the CSV files are 
the three feature vectors corresponding to
the embeddings of the radio image, the infrared image with radio contours, and the discussion text, respectively, for every radio subject.
In particular, the embeddings of radio images (columns \texttt{\small radio001} to \texttt{\small radio768}) and infrared images with radio contours (columns \texttt{\small ir001} to \texttt{\small ir768}) are produced by a pre-trained ViT model~\citep{dosovitskiy2020image}. The embeddings of the discussion text for radio subjects (columns \texttt{\small text001} to \texttt{\small text768}) are generated by a pre-trained BERT model~\citep{devlinBERTPretrainingDeep2019}.
Sections~\ref{sec:text_feature} and \ref{sec:image_feature} provide further details of how text and image features are extracted. 
%
%
The last 11 columns in the CSV files are binary indicators of tags for each radio subject.
%
The size of the dataset archive is 114 MB,
and its metadata is summarised in Table~\ref{tab:dataset}.

\begin{table}[!t]
\caption{
\daweicomment{Statistics of tags and text discussions of radio subjects}.
}
\label{tab:datasetbreakdown}
\resizebox{\textwidth}{!}{
\begin{tabular}{ r c c c c c c c } \toprule
\multicolumn{1}{c}{} & \textbf{Mean} & \textbf{Std Dev} & \textbf{Min} & \textbf{25\%} & \textbf{50\%} & \textbf{75\%} & \textbf{Max} \\ 
\midrule 
\textbf{\#Unique Tags} 
& 1.61 & 1.21 & 0 & 1 & 1 & 2 & 14 \\
\textbf{\#Comments} 
& 1.44 & 1.02 & 1 & 1 & 1 & 2 & 53 \\ 
\textbf{Word Count} & 10.96 & 11.99 & 1 & 4 & 8 & 14 & 424 \\ 
\bottomrule
\end{tabular}
}
\end{table}

\begin{table}[htbp]
\centering
\caption{
Metadata of the dataset archive on Zenodo.
}
\begin{tabular}{lr}
\toprule
Total Subjects & 10,643 \\
Subjects for Training & 7,688 \\ 
Subjects for Validation & 1,357 \\
Subjects for Test & 1,598 \\
Text Embedding Size & 768 \\
Radio Image Embedding Size & 768 \\
Infrared Image (with Radio Contours) Embedding Size & 768 \\
\bottomrule
\end{tabular}
\label{tab:dataset}
\end{table}


\section{CLASSIFIERS}
\label{sec:classifiers}
Building models capable of processing information from multiple modalities is essential for many applications within and outside astronomy~\citep{ngiam2011multimodal,baltruvsaitis2018multimodal,cuoco2021multimodal,hong2023photoredshift}, and in this work, we frame the task of predicting tags for radio subjects as a multi-label classification problem by employing a multi-modal machine learning approach to incorporate information from both text and image data in the RadioTalk dataset.

To extract text and image features we utilise state-of-the-art pre-trained transformer models. Transformers are a deep learning architecture that have demonstrated success for natural language processing and computer vision tasks, however require significant amounts of data to train \citep{lin2021survey}. Given our dataset is relatively small in the current era of deep learning we can leverage transfer learning by using pre-trained vision and language transformer models as feature extractors. 

We consider three sources of information for identifying the hashtag corresponding to each object:
1) radio image,
2) infrared image along with its contour lines,
and 3) free-form text from the RadioTalk discussion.
Each of these input data is converted to a numerical representation (called embedding) using deep learning. The image
data is embedded using a pre-trained Vision Transformer~\citep[ViT; ][]{dosovitskiy2020image}, and the text discussion is embedded using
a pre-trained language model called Bidirectional Encoder Representations from Transformers~\citep[BERT; ][]{devlinBERTPretrainingDeep2019}.

\subsection{Text features}
\label{sec:text_feature}

The task for our multi-label 
classifier is to predict tags associated with a given radio subject using the volunteer text comments. To extract a feature representation from the \eat{subject }comments, we performed three main steps. 

Firstly, we concatenated the comments for each radio subject, converted the text to lowercase, and removed all hashtags in the concatenated text. Secondly, we utilised the \texttt{spaCy} \citep[][]{spacy} library to perform text cleaning which included stopword removal and token lemmatisation.
A stopword is a word that occurs frequently but contributes little meaning, examples include `the', `is', and `for'. Lemmatisation is a process to reduce words to their root form, for example, the lemmatised form of `ejected' is `eject'. Both are common pre-processing techniques to remove noise in text.

We remark that while 
our
pre-processing of the free-form 
text were similar to that implemented by~\citet{bowles23}, our model does not predict for FR morphology classes from the tags. One difference between the text archive from~\citep{bowles23} and our work is that our RGZ volunteers have not been asked to describe the source morphologies in plain English within the RadioTalk forum, but rather to nominate source hashtags. Consequently, the RadioTalk forum contains fewer multi-tag and long text threads than that described in~\citep{bowles23}. The sparsity of our text data is also a driving reason for pursuing a multi-modal learning approach. 

Lastly, we perform feature extraction using 
a 
state-of-the-art pre-trained BERT model which has demonstrated success across a broad range of NLP tasks \citep{devlinBERTPretrainingDeep2019}.
There are many variants of the BERT transformer model, we used the \texttt{BERT-base-uncased}\footnote{\daweicomment{\url{https://huggingface.co/bert-base-uncased}}} model consisting of 12 transformer blocks~\citep{vaswani2017attention} and 110M parameters.
%
This pre-trained BERT model is powerful enough to handle long input context and is suitable for comments in the RadioTalk forum while being efficient in computation compared to much larger NLP models such as those adopted by~\citet{ciucua2023galactic}.
%
The model has a hidden size of 768, and thus a 768-dimensional embedding can be extracted for each token in the input text. However, we are interested in obtaining a single embedding for all text associated with each radio subject. BERT conveniently prepends a special \texttt{[CLS]} token to all input text, and the embedding of which is designed to use for classification tasks. 
%
%
The maximum number of input tokens (i.e. word pieces) for a text sequence specified by BERT is 512, and there are less than 0.7\% of radio subjects in our dataset are associated with text sequences longer than this limit. For simplicity, we truncated these long sequences before extracting the text features using BERT.

In this manner, we can extract a single 768-dimensional embedding for each radio subject in our dataset from the volunteer comments.

\subsection{Image Features}
\label{sec:image_feature}

\rgzcomment{Almost all of the information in the infrared image (>99\%) is irrelevant to the morphology tags assigned to the radio, and it seems implausible to me that training on the infrared images adds any value.  If you were training on the position in the infrared image identified by the RGZ users, or even a masked version of the IR image around the radio peaks, then I could see how value could possibly be added.}

Features of radio images and infrared images (with radio contours) were extracted by a pre-trained state-of-the-art ViT model.
ViT has recently been employed in galaxy morphological classification~\citep{lin2021galaxy} as well as detecting and segmenting objects in radio astronomical images~\citep{sortino2023radio}. It can learn positional embedding of image patches and is relatively efficient to train compared to classical convolutional neural networks\eat{ (CNNs)}~\citep{dosovitskiy2020image}.

In this work,
we used the \texttt{ViT-B/16}\footnote{\daweicomment{\url{https://download.pytorch.org/models/vit_b_16-c867db91.pth}}} architecture pre-trained on the ImageNet dataset~\citep{deng2009imagenet}. 
The radio images were first processed by a power stretch of 2 (i.e. squared stretch), and resized to 224x224 through linear interpolation of pixels.
They were then fed to the pre-trained ViT model
which split an input image into a sequence of patches of size 16x16, and the sequence of (flattened) patches were treated the same as tokens in a text sequence in BERT~\citep{dosovitskiy2020image}.

%
Similar to the embedding of the special \texttt{[CLS]} token in BERT, which serves as the sentence representation, ViT learns a embedding vector that serves as the representation of the sequence of image patches, in other words, a representation of the input image.
The features of infrared images with radio contours were extracted using the same approach, except that the preprocessing step only resizes the infrared images without performing any stretching.
As a result, we extract one 768-dimensional embedding for the radio image of a subject, and one 768-dimensional embedding for the infrared image of the same subject.
These image features can then serve as the input of a multi-label classifier to predict tags for a radio subject.

\subsection{Predicting hashtag probabilities}
\label{sec:predicting}

Let $\textbf{x}^{(i)}_\text{text}$, $\textbf{x}^{(i)}_\text{radio}$, $\textbf{x}^{(i)}_\text{ir}$ be the RadioTalk forum discussion text, the radio image, and the infrared image (with radio contours) of the $i$-th radio subject in the dataset, respectively.
We can create a numerical representation of the $i$-th radio subject using its text features as presented in Section~\ref{sec:text_feature} and/or its image features (Section~\ref{sec:image_feature}).
In particular, a representation of the $i$-th radio subject $\phi^{(i)}$ considered in this paper can incorporate text, image or multi-modal (i.e. text and image) information.
\begin{equation}
\label{eq:subject_feature}
\resizebox{.915\linewidth}{!}{$
    \phi^{(i)} = \begin{cases}
    \textsc{bert}(\textbf{x}^{(i)}_\text{text}), 
    & \text{\small text} \\
    f(\vit(\textbf{x}^{(i)}_\text{radio}), \, \vit(\textbf{x}^{(i)}_\text{ir})), 
    & \text{\small image} \\
    g(\textsc{bert}(\textbf{x}^{(i)}_\text{text}), 
    \vit(\textbf{x}^{(i)}_\text{radio}), \,
    \vit(\textbf{x}^{(i)}_\text{ir})),
    & \text{\small multi-modal} 
    \end{cases}
    $}
\end{equation}
for all $i = 1, \dots, N$ where $N$ is the number of radio subjects in the dataset.
$\textsc{bert}(\cdot)$ denotes features of the input text extracted by a pre-trained BERT model, 
and $\vit(\cdot)$ represents features of the input image extracted by a pre-trained vision transformer model. 
Functions $f$ and $g$ are feedforward neural networks (with learnable parameters) that combine multiple input embeddings into a single 768-dimensional embedding.

We then compute the probability $P_{ij}$ that the $j$-th tag is associated with the $i$-th radio subject in the dataset.
\begin{equation}
\label{eq:tag_proba}
    P_{ij} = \sigma(\mathbf{w}_j^\top \phi^{(i)}), \text{ for all $j = 1, \dots, 11$.}
\end{equation}
where $\sigma(z) = (1 + \exp(-z))^{-1}$ is the sigmoid function, and $\mathbf{w}_j$ is a 768-dimensional vector of weights which are learned by training the classifier.

A multi-label classifier compares $P_{ij}$ with a threshold probability $P_c$ (we used $P_c = 0.5$) and predicts the $j$-th tag is associated with the $i$-th radio subject if $P_{ij} \ge P_c$, otherwise it would suggests the $j$-th tag is not relevant to that radio subject.

\begin{figure}[t!]
\centering
\includegraphics[width=\linewidth]{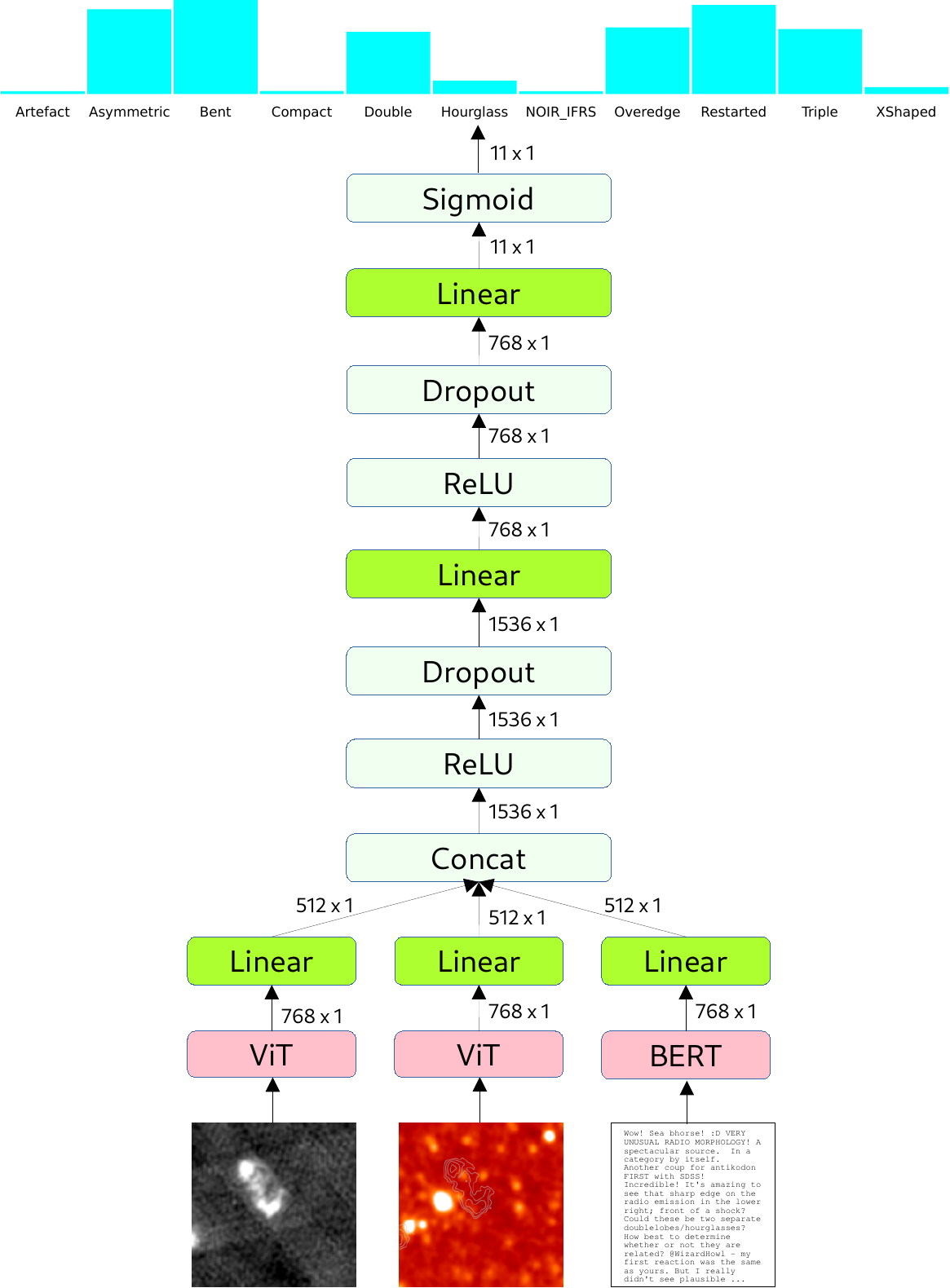}
\caption{
Architecture of the multi-modal classifier. 
As an example, this figure shows the predicted probabilities of the 11 tags for the radio subject ARG0002FUD, making use of its discussion text on the RadioTalk forum (after preprocessing as discussed in Section~\ref{sec:text_feature}), and its radio and infrared images.
}
\label{fig:clf_arch}
\end{figure}

\subsection{Multi-label classification methods}

There are many approaches to multi-label classification, but they can be broadly separated into two categories: problem transformation and algorithm adaptation \citep{tsoumakas2007multi}. Problem transformation methods approach the task by transforming the multi-label dataset into one or more single-label datasets. From here generic machine learning approaches can be employed to train a classification model per label. On the other hand, algorithm adaptation approaches make fundamental changes to the training and prediction mechanism to directly handle multiple labels simultaneously \citep{Bogatinovski2022}. In our work we employ two problem transformation approaches, binary relevance and classifier chains~\citep{read2009classifier,clf_chains_paper}.

Binary relevance is a simple approach to multi-label classification, which involves training a binary classifier for each label. 
Multi-label classifiers based on Equation~\eqref{eq:tag_proba} are binary relevance classifiers. 
In this paper, we refer to the multi-label classifier based solely on the text features as the \textit{text classifier}, and the one that uses features of the radio and infrared images but not the text features as the \textit{image classifier}. 
The multi-label classifier that makes use of multi-modal information (i.e. text and image features) 
is denoted as the \textit{multi-modal classifier}.

Figure~\ref{fig:clf_arch} illustrates the architecture of the multi-modal classifier used in this work. 
Given the RadioTalk forum text, the radio and infrared images of a radio subject,
the multi-modal classifier first combines the text features extracted by BERT and features of both the radio and infrared images extracted by ViT, and then passes the combined representation to an MLP to produce the probabilities which indicate how likely each of the 11 tags will be labelled for the radio subject.

A weakness of the binary relevance approach is that it cannot capture dependencies between labels.
To extend it to handle label dependencies we employed classifier chains. Each classifier, except the first in the chain uses the input features and predictions from the previous classifier to predict the target label. Again any base binary classifier can be used however performance will be sensitive to the ordering of labels in the chain.
In this work, we followed the recommendations by~\cite{read2009classifier} and averaged an ensemble of 10 randomly ordered chains when making predictions.

\subsection{Classification performance evaluation}

To evaluate the performance of our classifiers we used standard metrics including Precision, Recall, F1-score and Balanced Accuracy for each label. 
We can summarise the performance across labels by taking the macro-average. We also used a few ranking metrics specific to the task of multi-label classification, including Label Ranking Average Precision (LRAP), Coverage Error and Label Ranking Loss \citep{tsoumakas2009mining}.
\daweicomment{More details of these performance metrics can be found in~\ref{appendix:metrics}.}
We remark that multi-label classification is a significantly more difficult task than multi-class classification, since the number of possible label sets grows quickly (specifically, it grows by $n\choose{k}$ where $n$ is the total number of classes, and $k$ is the maximum number of allowed labels), a random classifier would return an F1 score very close to zero.

\subsection{Multi-label prediction and open set classification}
\label{sec:multilabel_vs_openset}

Radio galaxy classification is an important task in astronomy, 
and we remark that, in machine learning context, the word ``classification'' unfortunately has a very specific meaning which is usually different from the meaning of classification in astronomy.
In particular, multi-class classification typically assumes that all classes are known, and that each data point
has one and only one possible class, while multi-label classification allows multiple classes (from a known set of classes) for a data point. 
There are two desiderata for classifying radio morphologies: 1) identification of new morphologies, and 2) identifying multiple possible descriptions for a particular radio subject.
Therefore, if we want \daweidel{to allow }multiple hashtags per radio subject (as indicated by the RadioTalk forum data), 
then \daweidel{we need }multi-label classification is needed since the set of tags (i.e. classes) are known in our dataset. 
On the other hand, if one is interested in  identifying new morphologies, it is more appropriate to use open set classification~\citep{scheirer13} which does not assume all classes are known, and we leave this for future work.

\section{RESULTS}
\label{sec:res}

\subsection{Over 10,000 new sources beyond the DR1 catalogue}
\label{subsec:complexsources}

\begin{figure}[!h]
  \centering
  \includegraphics[width=0.58\textwidth]{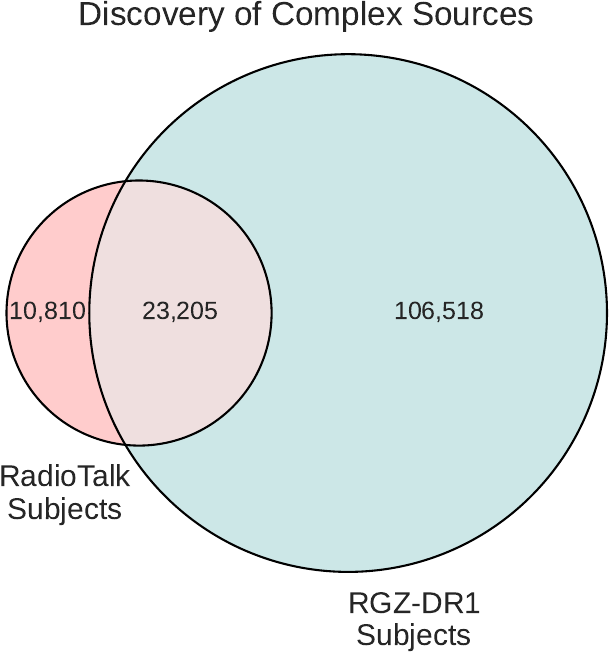}
  \caption{Over 10,000 complex sources \daweicomment{that are not in the RGZ-DR1 catalogue} are available in the RadioTalk dataset.
  }
  \label{fig:venn_diagram} 
\end{figure}

All subjects in the RGZ project were assigned unique identifiers shared across the forum and the catalogue. The catalogue contains the astronomical meta-data and the volunteer generated annotations associated with each radio source in a subject. We expected the forum subjects to be a subset of the catalogue for which users provided tags or comments. However, we found an interesting result when we attempted to cross-reference subjects in the forum with those in the catalogue using their common identifier. We discovered $10,810$ unique subjects that were present in the forum but not in the catalogue (see Figure~\ref{fig:venn_diagram}). Given each RGZ subject required $20$ independent classifications, we recovered over $200,000$ volunteer classifications. This finding implied volunteers could interact with these subjects, but their collective annotation results were not reaching the catalogue.

After investigating these subjects manually, we found they were extended, complex and noisy sources and thus would have been difficult to classify. 
The RGZ pipeline uses the Kernel Density Estimator (KDE) method\footnote{KDE is a kernel-based method to estimate the probability density function. Within the RGZ pipeline, KDE is applied to the volunteers' clicks.} to converge on the source classification from the many volunteers' classifications \citep{banfieldRadioGalaxyZoo2015}.   For complex subjects, the classifications from the many volunteers may contradict one another.  This can result in KDE failing to converge, in addition to low classification consensus levels below the threshold required for inclusion in RGZ DR1 \citep{wong23}.  Given the disagreement amongst volunteers, these complex sources are likely to contain unique or interesting qualities which are ripe for further analysis.

\begin{table*}[!ht]
\centering
\caption{Classification performance (in terms of F1 score) of each tag, for text, image, and multi-modal information. Classes are sorted in increasing order of the number of subjects tagged with the class. 
\textbf{Text}: Binary relevance classifier using only text features; \textbf{Text (CC)}: Classifier chains using only text data; \textbf{Image}: Classifier using only images; \textbf{Text+Image}: Classifier using multi-modal information.
The best performance for each tag is shown in \best{bold italic}.
}
\begin{tabular}{lp{1cm}p{1cm}p{1cm}p{1cm}p{1cm}p{1cm}p{1.2cm}p{1cm}p{1cm}p{1cm}p{1cm}p{1cm}}
\toprule
       & xshaped & restarted & artefact & noir\_ifrs & bent & overedge & asymmetric & hourglass & triple & compact & double \\
\midrule
Text+Image & \best{0.158} & 0.176 & \best{0.109} & \best{0.385} & \best{0.364} & \best{0.511} & \best{0.257} & 0.373 & \best{0.423} & 0.643 & 0.637  \\
Image  & 0.118 & \best{0.178} & 0.104 & 0.325 & 0.347 & 0.455 & 0.256 & \best{0.377} & 0.403 & \best{0.648} & \best{0.641}  \\
Text  & 0.021 & 0.053 & 0.076 & 0.331 & 0.209 & 0.419 & 0.236 & 0.228 & 0.313 & 0.347 & 0.473 \\
Text (CC)  & 0.011 & 0.039 & 0.027 & 0.242 & 0.150 & 0.231 & 0.208 & 0.209 & 0.273 & 0.380 & 0.510 \\
\bottomrule
\end{tabular}
\label{tab:f1-compare-text-image-hybrid}
\end{table*}

\begin{figure*}
    \centering
    \includegraphics[width=\textwidth]{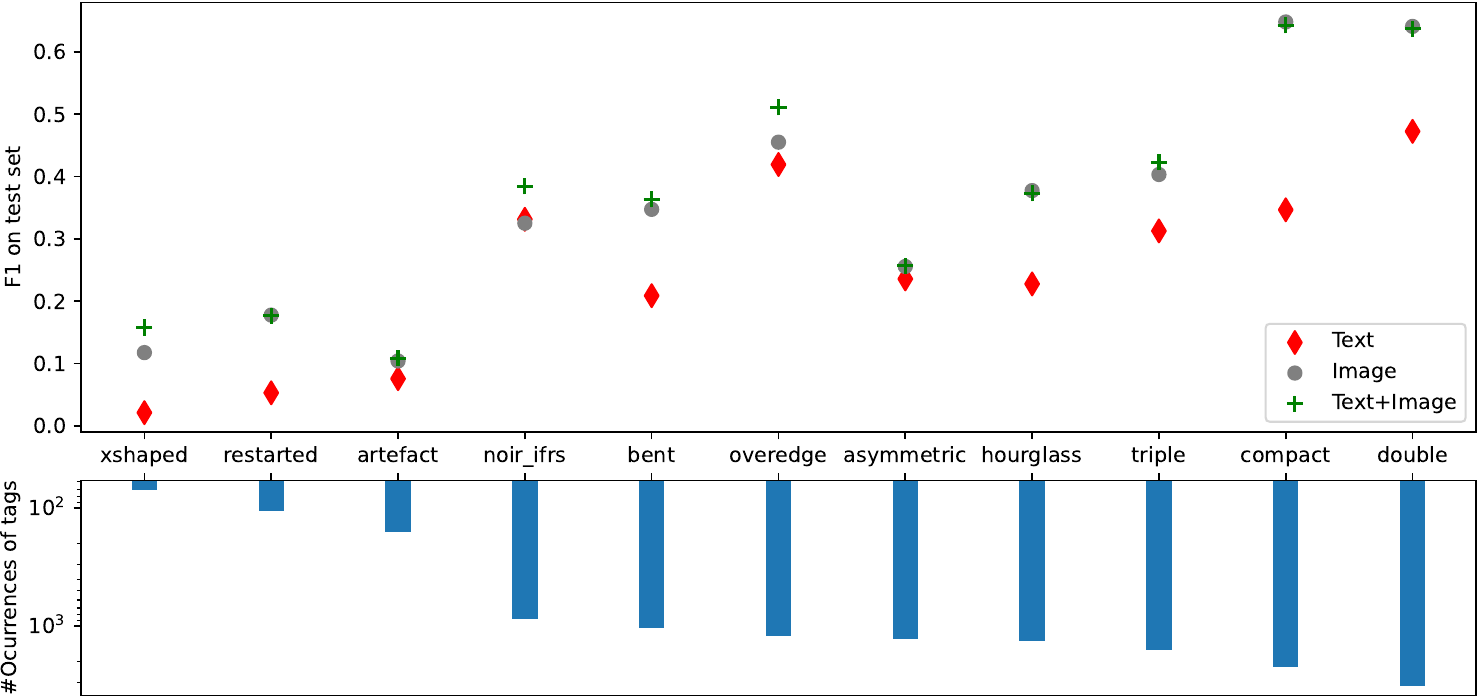}
    \caption{
    Classification performance (in terms of F1 score) versus the number of occurrences of tags in the dataset for the binary relevance classifiers using text (\textbf{Text}), images (\textbf{Image}) and multi-modal information (\textbf{Text+Image}).
    Text information can be helpful in predicting tags with few (e.g. \texttt{xshaped}), medium (e.g. \texttt{noir\_ifrs} and \texttt{overedge}) or large (e.g. \texttt{triple}) number of observations in the dataset.
    }
    \label{fig:f1_vs_tags}
\end{figure*}

\subsection{Text information complements image data}

When text and image features individually result in good and comparable accuracy, the multi-modal features give significantly better performance 
(see Table~\ref{tab:f1-compare-text-image-hybrid}).
For example, the individual text and image classifiers were performing relatively well in the \texttt{noir\_ifrs} and \texttt{overedge} classes.  However, this performance is further boosted in the multi-modal classifier when both sets of information were used in combination.

This result suggests that the text conversations between the citizen scientists and academic scientists are contributing additional information to the radio morphologies that are not as evident in the radio and IR heatmap images.

Specifically, there are two main avenues for additional information that the text data have relative to the image data: 1) participants are more likely to discuss and tag complex sources which they find interesting or are confused by; and 2) all participants in the RadioTalk forum are provided with additional tools to zoom in and out of a subject, in addition to links to coordinate-matched sky images from other multiwavelength surveys. Such additional tools are useful for providing additional information about a subject that otherwise would not be possible with the heatmap images alone.  

Hence, through the use of a multi-modal classifier, we are able to take full advantage of all the information that is present within RGZ.  Such an approach contributes towards maximising the scientific value that could be gained from the investment of effort in conducting a citizen science project.
Interestingly, when using only text data, we found that classifier chains barely improve upon the binary relevance classifier across the standard metrics used to evaluate the performance for most tags (see Table~\ref{tab:f1-compare-text-image-hybrid}),
in addition to the macro-average metrics as well as ranking metrics such as LRAP, with recall as an exception (see Tables~\ref{tab:clf_dr1} and \ref{tab:clf_dr1_tag} in~\ref{appendix:results}).

\subsection{Rare morphologies are harder to classify}

As expected, classes with larger sample sizes results in better performance on average (see Table~\ref{tab:f1-compare-text-image-hybrid} and Figure~\ref{fig:f1_vs_tags}).
However, diverse or more complex morphologies (such as `asymmetric') result in lower accuracy even when there is a large sample size within its class (as shown in Figure~\ref{fig:f1_vs_tags}).

\section{LIMITATIONS AND IMPLICATIONS}

\rgzcomment{I downloaded the training set, and see some problems with the reliability of the classifications.  Not surprising, but question remains how these reliability problems affect the training, etc.   For example, of the 1592 compact sources, 347 have another classification (other than noir).  Of the 848 overedge sources, 732 of them have another classification (other than noir).  Of the 121 artefacts, 104 of them have another classification.  All of these are incompatible combinations.}

The limitations inherent in our current study comes from two main sources: 1) limitations from data collection that is carried forward from the RGZ project to our study here; 2) the methods that we use to investigate the relative efficacy of multi-modal classification.

\subsection{Limitations of the RGZ dataset}

The RGZ project is based on the publicly-available observations of the FIRST survey \citep{white97}.  In comparison to contemporary radio surveys such as the Rapid ASKAP Continuum Survey \citep[RACS; ][]{hale2021racs}, the FIRST survey is much less sensitive to diffuse and extended emission.  Therefore, the performance of our image and 
multi-modal classifiers will also be limited by the intrinsic imaging limitations that comes from the FIRST survey.

Annotations and tags from citizen science projects can be subjected to a range of potential biases \citep[e.g.\ ][]{draws2021}.  For example, the motivation for discovering new objects may lead to potential biases in the classes of biases such as confirmation and anchoring biases.  However, our results in this paper (see Section~\ref{sec:res}) suggest that our tag dataset is sufficiently accurate for this study.  As such our participants' personal motivation for discovery is unlikely to lead to tags that are strongly influenced by confirmation or anchoring biases.

As mentioned previously, the task of tagging is optional within the RGZ workflow which results in both data sparseness and class imbalance.  To reduce the impact of such limitations, we merged similar classes 
(see Table~\ref{tab:merge_tags}) to increase the number of samples for the classes being studied.

The complexity of radio galaxy morphologies depends on the observational limitations of the instruments that are used. Hence, the relative importance of tagged morphologies may evolve in time. For example, observations at higher resolution and sensitivity of the same source that was previously classified may reveal more detailed morphologies which could change the classification. Therefore, it is possible that the tags that are derived from the RGZ project may not be as useful for future next-generation surveys.

\subsection{Limitations of our results}

The tagging problem investigated in this work is formulated as a multi-label classification problem, which assumes that all classes (that is, tags) are known before tagging a radio subject. However, the set of tags we used obviously cannot characterise every radio subject in the universe, this implies that if new morphologies are needed, for example, in another dataset, the proposed approach will not accurately predict the tags. In this case, a different approach such as open set classification is needed (see the discussion in Section~\ref{sec:multilabel_vs_openset}).

Another limitation of our results comes from the  dramatic differences in the frequencies of tags (see Figure~\ref{fig:tag_usage}), even after we merge similar tags, as shown in Figure 6. Such imbalanced class sizes may result in less accurate classification performance.

The text and image features used in the multi-modal classifier are extracted by pre-trained deep learning models that are not specifically trained to classify tags for radio subjects. Using model weights further optimised for our multi-label classification problem (e.g. through fine-tuning those pre-trained deep models on the RadioTalk dataset) or employing deep models specifically designed for our problem may lead to more accurate classification of tags.


\subsection{Implications for future citizen science projects}

Thanks to advances in radio telescope technologies, astronomers will be able to survey tens of millions of radio sources \citep[e.g.\ ][]{norris21} but without the ability to visually classify even a few percent of such samples. Hence, citizen science projects such as RGZ provide the potential for astronomers to obtain quantified visual classifications in a more efficient manner.  Furthermore, the accuracy and fidelity of such citizen science classifications have been further demonstrated by the success of early machine learning-based prototype classifiers which used pre-release versions of RGZ-DR1. 

In this paper, we demonstrate the value of multi-modal (text+image) classification.  
Using the \texttt{overedge} class as an example, 
we show greater F1 scores than if we were to base our classifier solely on the images or the text data alone (see Figure~\ref{fig:f1_vs_tags}).
%
\daweicomment{This indicates that both text and image information add value in terms of classification performance.
}

As the next-generation radio surveys get underway to reveal new classes of radio sources with morphologies that were not seen before \citep[e.g.\ ][]{norris2021orcs}, the relative importance of tagged morphologies will continue to evolve.  Based on the premise that new discoveries may not fit well into historically-derived categories of sources and paradigms, \citet{rudnick2021} argues for the importance of tags for the classification of sources for future surveys.  Building on the arguments of \citet{rudnick2021}, our results provide quantitative and empirical support for the use of hashtags and text, in combination with images in future citizen science projects for the classification of radio galaxies.  Indeed, this is a recent topic of active research and our study here can also be compared to recent work by \citet{bowles23} that delve into improving the current classification scheme through developing
\daweicomment{a semantic taxonomy for radio morphology.}

In the future, automated classifiers based on machine learning methods will continue to improve and reduce the fraction of sources that require visual inspection. However, advances in instrumentation will nevertheless probe new parameter spaces thereby retaining the need for visual verification.  Therefore, citizen science projects (which gather 
both images and text conversations) will continue to be important for such at-scale exploration of the Universe.

\section{SUMMARY AND CONCLUSIONS}

As the next-generation SKA pathfinder surveys embark on large sky surveys that produce tens of millions of radio galaxies, we do not have sufficient scientists (including citizen scientists) to perform visual inspection and classification of such large samples.  Therefore, the development of machine learning-based automated classifiers is crucial for increasing the discovery rate of rarer and more complex classes of sources by citizen scientists and the academic science team \citep[e.g.\ ][]{gupta22}.
Participants in RGZ were afforded the option to tag or label subjects with hashtags in addition to contributing towards further discussion either on the specific subject or on any broader science topic that is related to the phenomena such as 
radio galaxy evolution.

This study investigated the use of machine learning on free-form text discussion, to complement multiwavelength images of radio galaxies,
for the task of automating the identification of galaxy morphologies.
We explored the use of multi-modal classification to determine whether the addition of hashtag labels will result in improved accuracy with classifying rare radio galaxy morphologies.
In a one-to-one comparison, we found that the image classifier outperforms the text classifier across most performance metrics (see Table~\ref{tab:clf_dr1} in~\ref{appendix:results}).  The combined text and image classifier results in improved performance over the image classifier and text classifier alone.  Such a result suggests that future RGZ-like citizen science projects may benefit from the increased use of hashtags in order to build more accurate models of the different radio galaxy classes.
Unsurprisingly, rare morphologies are not well represented in the data. We used a hierarchical tree (see Figure~\ref{fig:tagtree}) to merge the smaller classes to improve the sample sizes for the rarer morphology classes.  While the reduction of classes does result in a small improvement in F1 scores, due to the small sample sizes of the various morphological classes, we recommend caution
in interpreting the performance metrics for our classification methods.

We find that pre-trained language models (like BERT) and vision transformer have significant potential for the task of classifying complex radio source morphologies. In particular, we show the potential of quantifying 
free-form text in astronomy, using deep learning models that represent text as vectors of numerical features.
Text is becoming increasingly important in citizen science \citep{rudnick2021,bowles22,bowles23}, and our results
provide further evidence supporting the use of free-form discussions and hashtags, 
to complement image analysis for galaxy classification. However, our results do suggest that future citizen science projects should address the fact that rarer classes of objects will result in fewer tags which could lead to class imbalance issues.  This proof-of-concept paper has demonstrated the use of data driven approaches to improve our taxonomy of radio galaxies, as well as the
benefits of richer multi-modal sources of information using machine learning based methods.

\begin{acknowledgement}
We acknowledge the traditional custodians of the lands upon which the work and writing of this paper was completed. This includes the land of the Ngunnawal and Ngambri people, and the land of the Whadjuk Noongar people. We thank our anonymous referee and associate editor for their constructive suggestions which further improved the clarity of this paper.

This publication has been made possible by the participation of more than 12,000 volunteers in the Radio Galaxy Zoo project. Their contributions are individually acknowledged at 
\texttt{\small{\url{http://rgzauthors.galaxyzoo.org}}}.
\end{acknowledgement}



\bibliography{references}

\appendix
\section{The RGZ interface}
\label{appendix:rgzinterface}

\begin{figure}[htbp]
\centering
\includegraphics[height=8in]{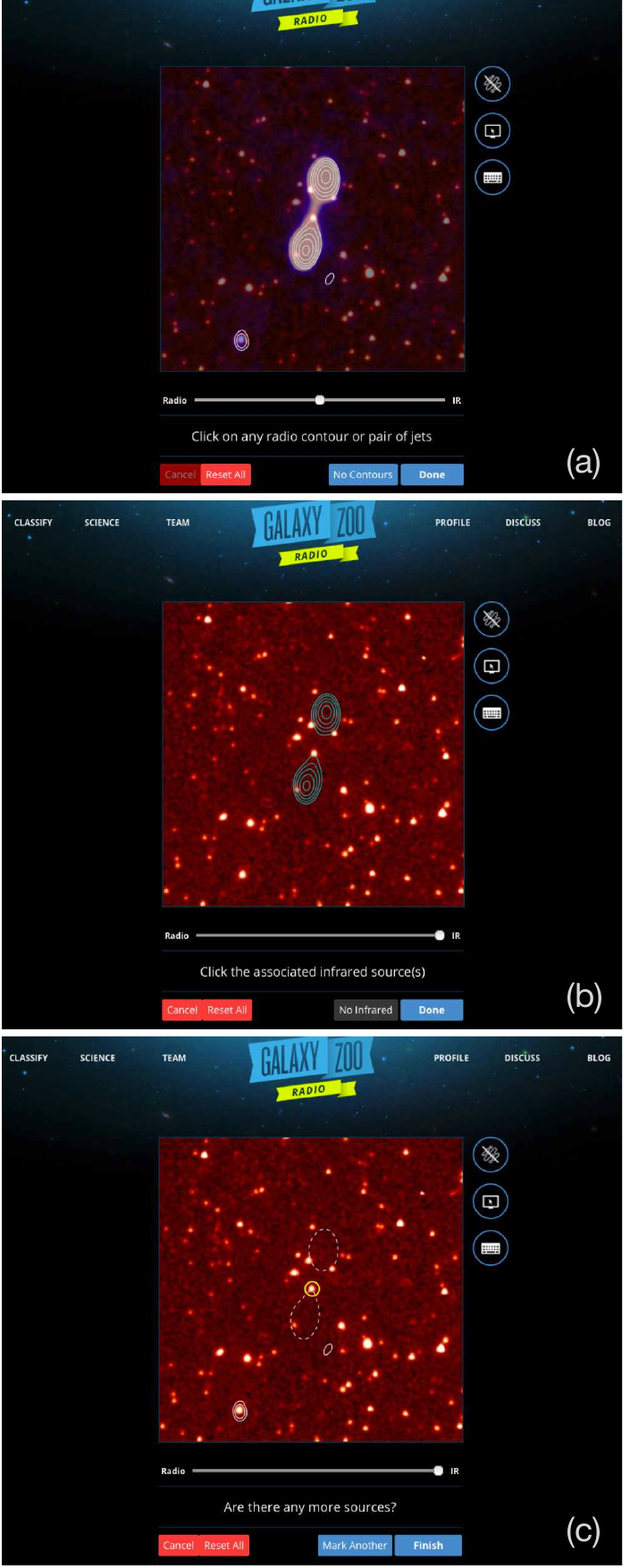}
\caption{RGZ interface for cross-matching radio components to host galaxies \citep{banfieldRadioGalaxyZoo2015}. Panel (a) shows an example double-lobed radio source and the slider in the central position where both the radio and infrared images are presented in blue and orange heatmaps, respectively. As the slider is transitioned completely towards IR, the radio image reaches 100~percent transparency and the radio emission is represented by the sets of contours (panels b and c). The associated radio components are highlighted as blue contours in panel (b) and the volunteer-identified cross-matched host galaxy is marked by the yellow circle in panel (c).}
\label{fig:rgz_interface}
\end{figure}

The RGZ online interface is shown in Figure~\ref{fig:rgz_interface}.

\section{Examples of radio subjects}
\label{appendix:examples}
\begin{figure*}[htbp]
    \setlength{\tabcolsep}{4pt}
    \begin{tabular}{ccccccccccc}
    \toprule
    \texttt{artefact} & \texttt{asymmetric} & \texttt{bent} & \texttt{compact} & \texttt{double} & \texttt{hourglass} & \texttt{noir\_ifrs} & \texttt{overedge} & \texttt{restarted} & \texttt{triple} & \texttt{xshaped} \vspace{.3em} \\
    \includegraphics[height=0.53in]{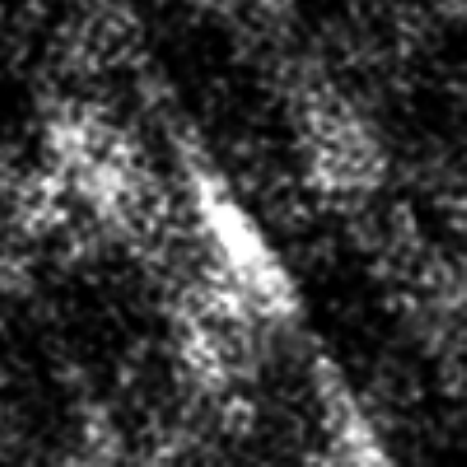} &
    \includegraphics[height=0.53in]{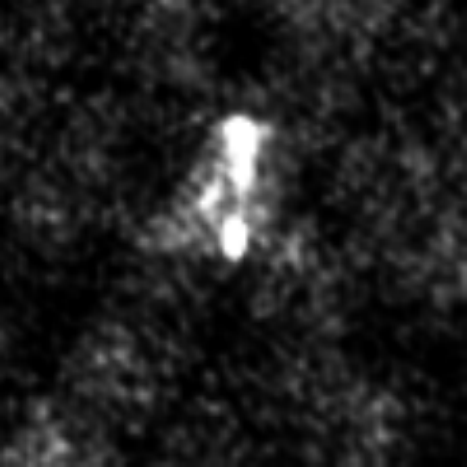} &
    \includegraphics[height=0.53in]{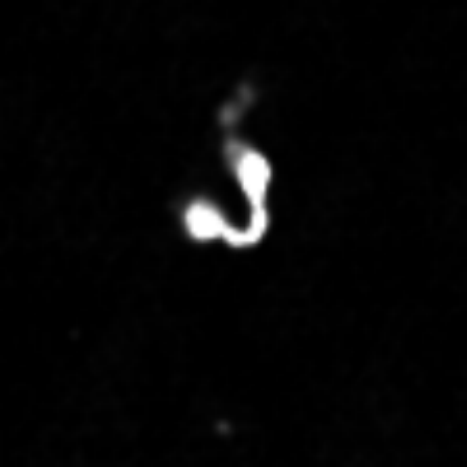} &
    \includegraphics[height=0.53in]{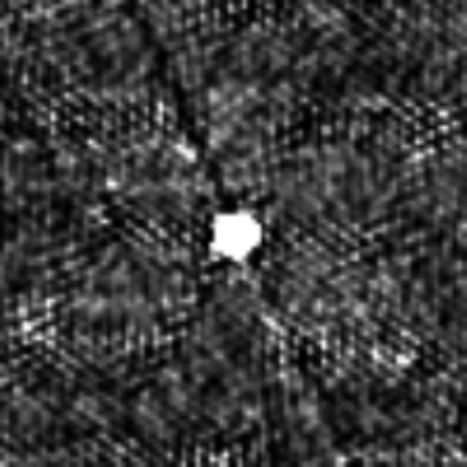} &
    \includegraphics[height=0.53in]{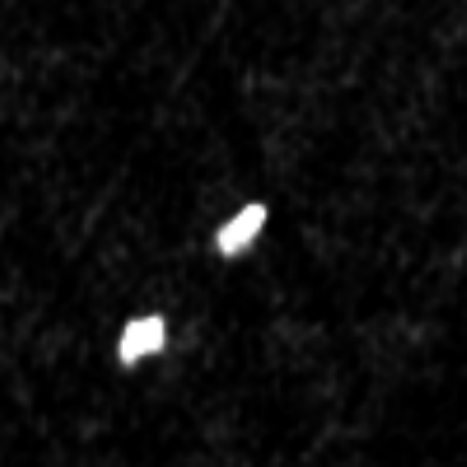} &
    \includegraphics[height=0.53in]{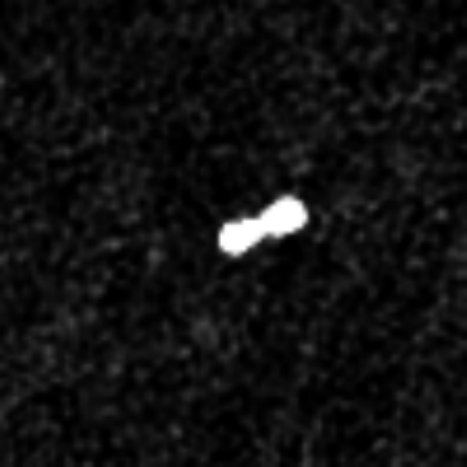} &
    \includegraphics[height=0.53in]{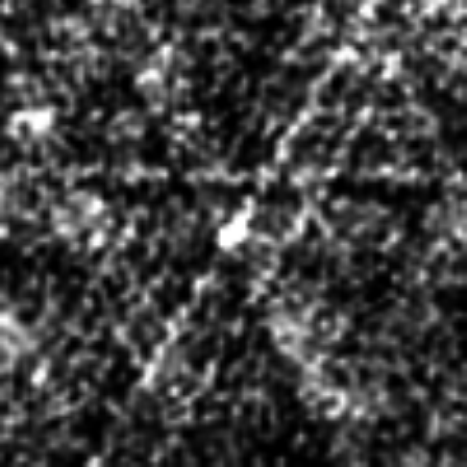} &
    \includegraphics[height=0.53in]{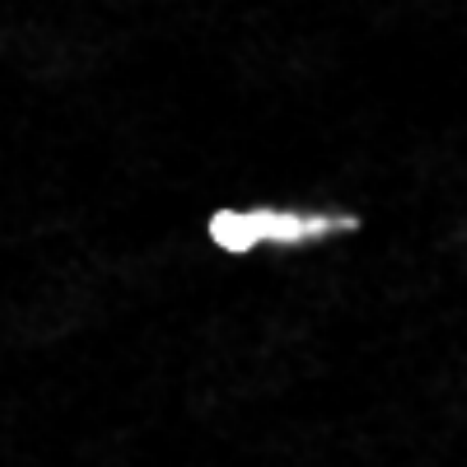} &
    \includegraphics[height=0.53in]{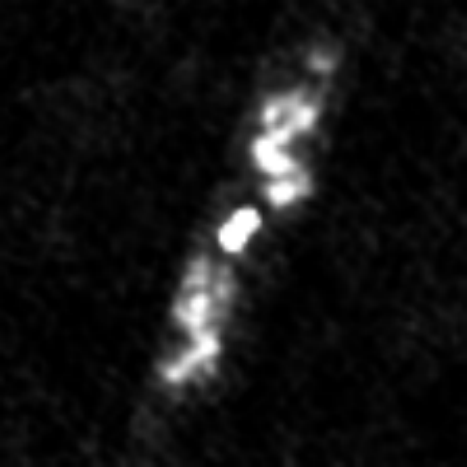} &
    \includegraphics[height=0.53in]{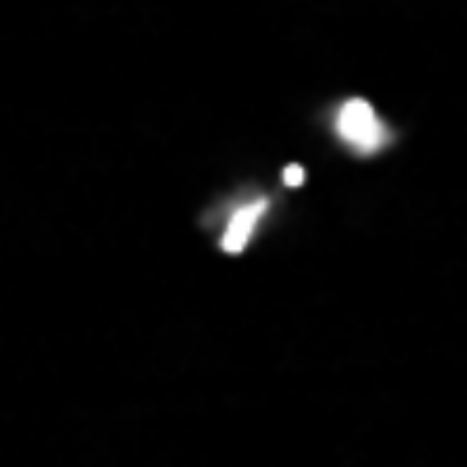} &
    \includegraphics[height=0.53in]{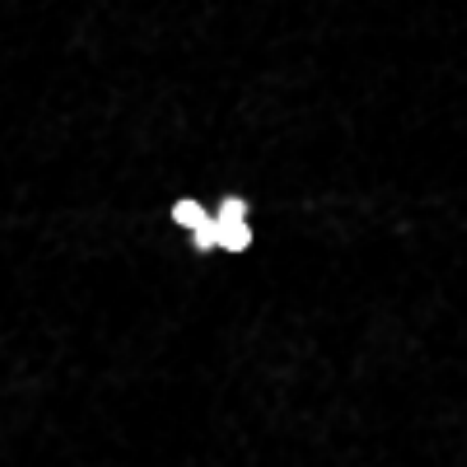} \\
    \bottomrule
    \end{tabular}
    \caption{Examples of radio subjects for each of the 11 tags. From left to right: 1) \texttt{artefact} describes an image where there is significant image processing residuals; 2) \texttt{asymmetric} describes radio jets or lobes that are not symmetrical; 3) \texttt{bent} describes jets and lobes that appear to have been swept to one side; 4) \texttt{compact} describes an unresolved single component radio source; 5) \texttt{double} describes two radio components that extend away from the host galaxy; 6) \texttt{hourglass} describes two overlapping radio components; 7) \texttt{noir\_ifrs} describes a radio source with no visible galaxy counterpart; 8) \texttt{overedge} describes a source which extends beyond the field-of-view of the image; 9) \texttt{restarted} describes a radio jet that consists of more than 3 components which could be due to restarting radio jet activity; 10) \texttt{triple} describes a source with 3 radio components; 11) \texttt{xshaped} describes a source which appears to be a superposition of 2 orthogonal hourglass sources.}
    \label{fig:tag_example}
\end{figure*}

Figure~\ref{fig:tag_example} shows examples of radio subjects for the 11 tags used in this work.

\section{Background on performance metrics for multi-label classification}
\label{appendix:metrics}

The classification performance is evaluated using Precision, Recall and F1 score for each tag (i.e. label), as well as their macro-averaged counterparts over all tags. In particular, let $N$ be the number of samples (i.e. radio subjects) in the dataset, $M$ be the number of tags, and $TP_j$, $FP_j$, $TN_j$ and $FN_j$ be the number of true positives, false positives, true negatives and false negatives for the $j$-th label, respectively. 
The Precision, Recall and F1 score for the $j$-th label are defined as
\[
\begin{aligned}
&\text{Precision}_j = \frac{TP_j}{TP_j + FP_j}, \\
&\text{Recall}_j = \frac{TP_j}{TP_j + FN_j}, \\
&\text{F1}_j = \frac{1}{\tfrac{1}{2} \left( \tfrac{1}{\text{Precision}_j} + \tfrac{1}{\text{Recall}_j} \right)},
\end{aligned}
\]
and the macro-averaged definitions are
\[
\begin{aligned}
&\text{macro-Precision} = \frac{1}{M} \sum_{j=1}^M \text{Precision}_j, \\
&\text{macro-Recall} = \frac{1}{M} \sum_{j=1}^M \text{Recall}_j, \\
&\text{macro-F1} = \frac{1}{M} \sum_{j=1}^M \text{F1}_j.
\end{aligned}
\]

Further, given that the dataset is imbalanced (as discussed in Section~\ref{sec:dataset}), we also report the performance using the balanced accuracy~\citep{brodersen2010balanced,kelleher2015fundamentals} averaged over all tags, i.e.
\[
\text{Balanced Accuracy} = \frac{1}{2M} \sum_{j=1}^M \left( \frac{TP_j}{TP_j + FN_j} + \frac{TN_j}{TN_j + FP_j} \right).
\]

Lastly, we report the performance of the multi-label classification using three ranking metrics: Label ranking average precision (LRAP), coverage error and label ranking loss. Recall that the multi-label classifier gave a probability for every tag for a given radio subject (as detailed in Section~\ref{sec:predicting}). Let $Y \in \{0, 1\}^{N \times M}$ be the binary indicator matrix of true labels (that is, $Y_{ij} = 1$ iff the $i$-th radio subject was given the $j$-th tag), and $rank_{ij}$ be the rank of the $j$-th tag given by the multi-label classifier (in terms of predicted probabilities). The three ranking metrics are defined as
\begin{align*}
&\text{LRAP} = \frac{1}{N} \sum_{i=1}^N \frac{1}{m_i} \sum_{\substack{\ \ j \in \{1,\dots,M\} \\ Y_{ij} = 1}} \frac{m_{ij}}{rank_{ij}}, \\
&\text{Coverage Error} = \frac{1}{N} \sum_{i=1}^N \max_{\substack{\ \ j \in \{1,\dots,M\} \\ Y_{ij} = 1}} rank_{ij}, \\
&\text{Label Ranking Loss} = \frac{1}{N} \sum_{i=1}^N \frac{q_i}{m_i (M - m_i)},
\end{align*}
where $m_i$ is the number of ground-truth tags of the $i$-th radio subject, 
$m_{ij}$ is the number of ground-truth tags with higher probabilities (as predicted by the multi-label classifier) than that of the $j$-th tag,
and $q_i$ is the number of tag pairs that are incorrectly ordered by the multi-label classifier (i.e. a ground-truth tag is given a lower probability than that of a false tag).

\section{More detailed results}
\label{appendix:results}

\begin{figure*}[htbp]
\centering
\setlength{\tabcolsep}{.1in}
\vspace{0.14in}
\begin{tabular}{ccccc}
\includegraphics[height=1.35in]{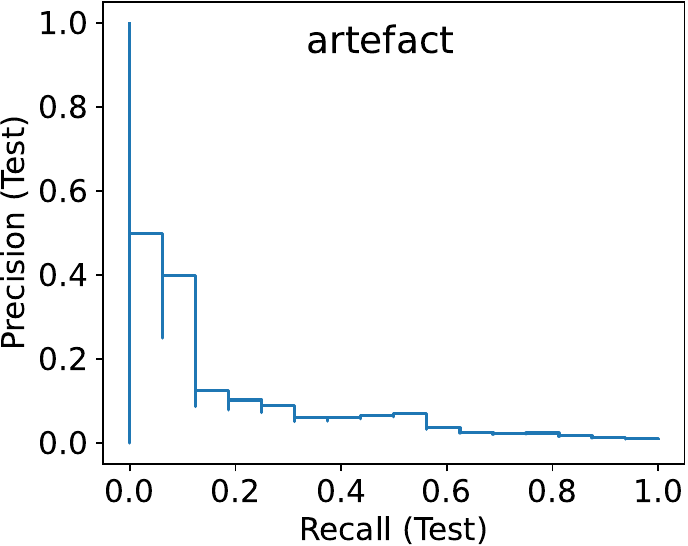} & {} &
\includegraphics[height=1.35in]{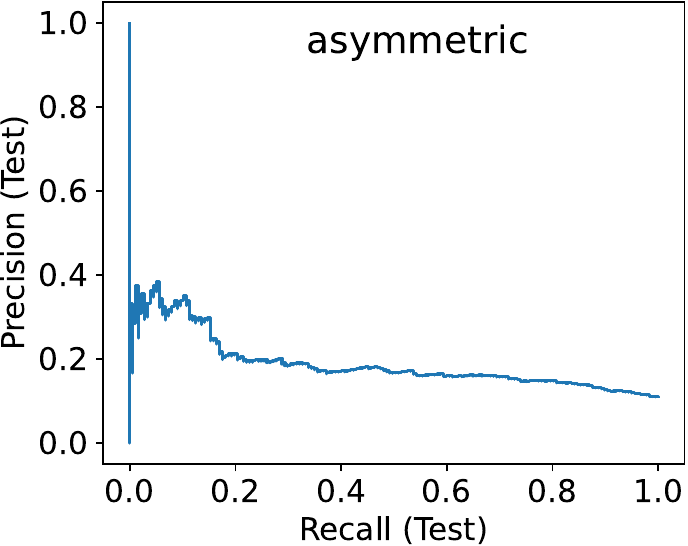} & {} &
\includegraphics[height=1.35in]{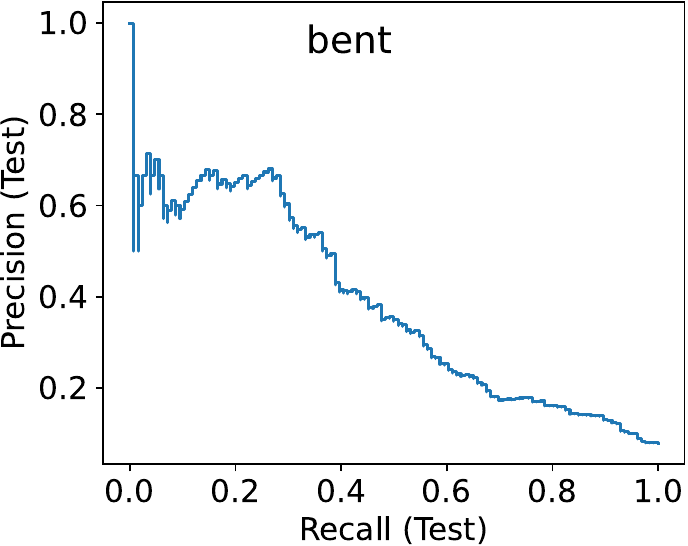} \vspace{0.1in} \\
\includegraphics[height=1.35in]{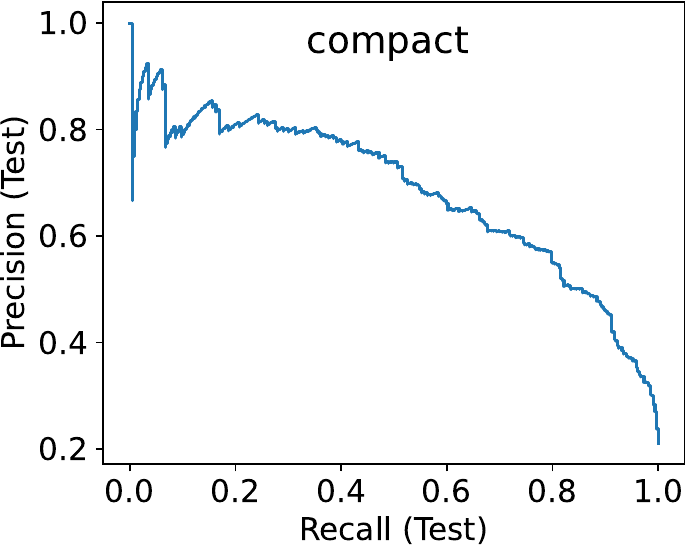} & {} &
\includegraphics[height=1.35in]{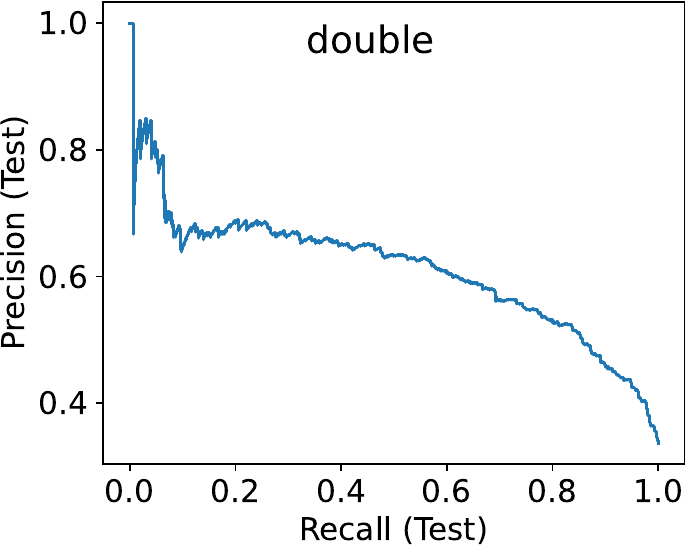} & {} &
\includegraphics[height=1.35in]{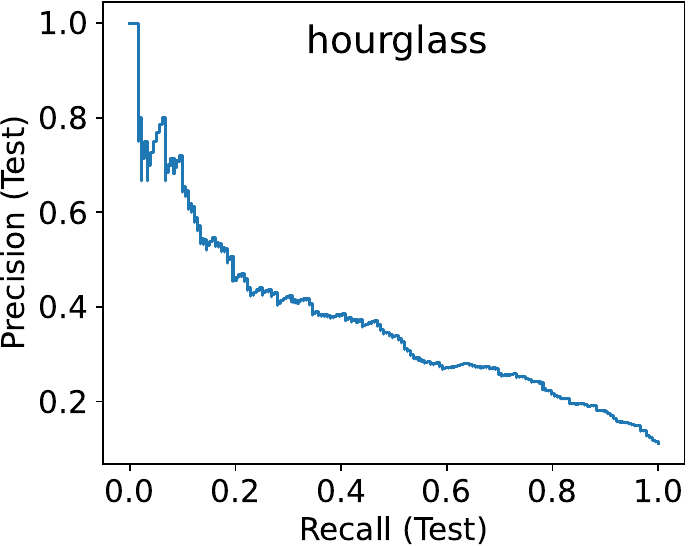} \vspace{0.1in} \\
\includegraphics[height=1.35in]{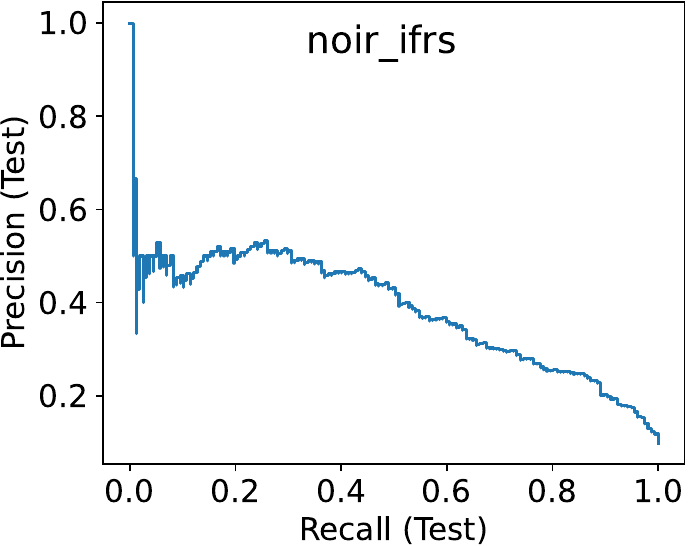} & {} &
\includegraphics[height=1.35in]{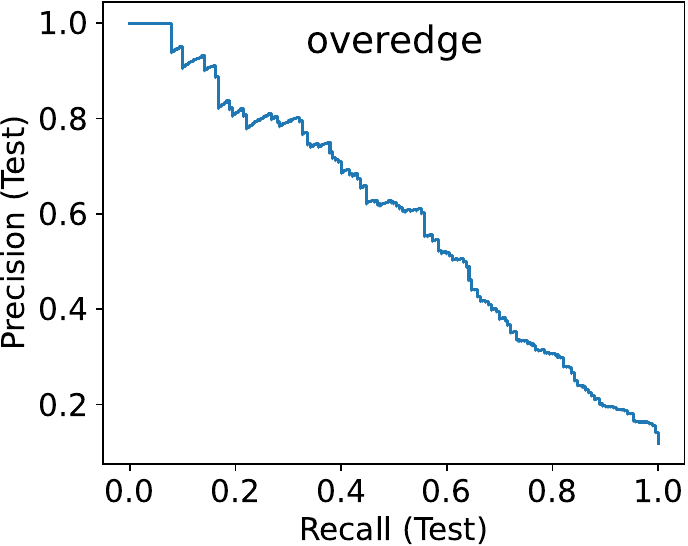} & {} &
\includegraphics[height=1.35in]{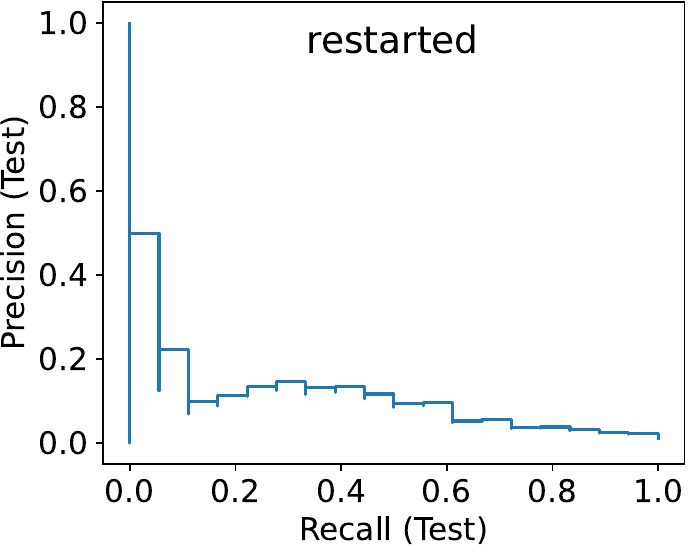} \vspace{0.1in} \\
\includegraphics[height=1.35in]{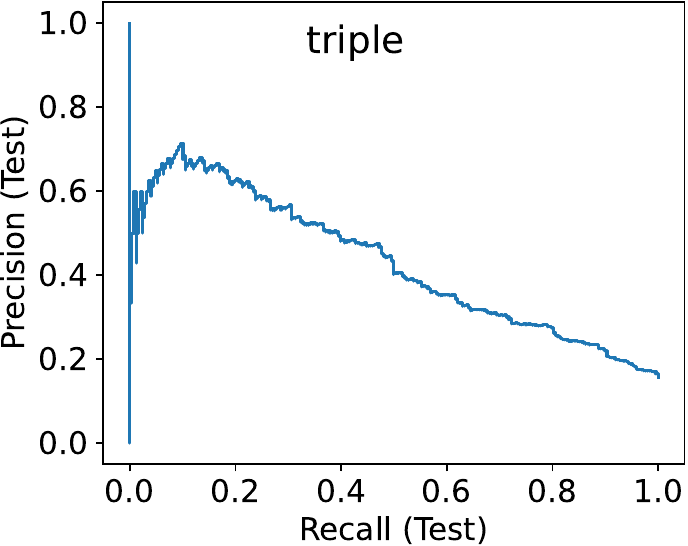} & {} &
\includegraphics[height=1.35in]{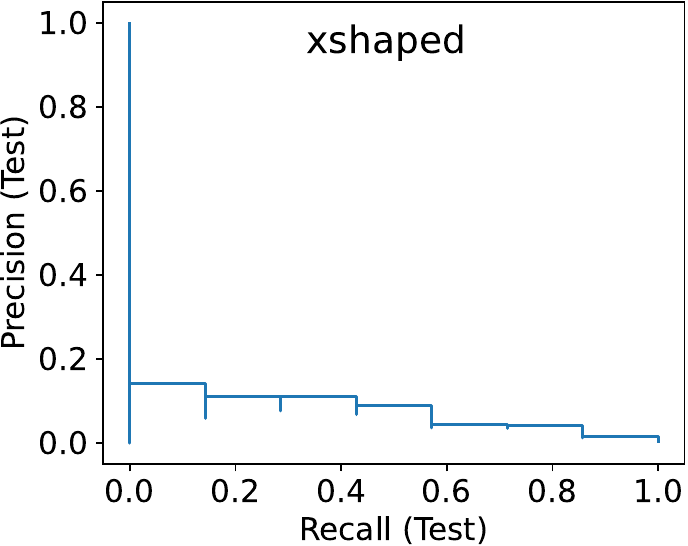} & {} & {}
\end{tabular}
\caption{%
Precision-Recall curves of the multi-label classifier using multi-modal information for each of the 11 tags.
} 
\label{fig:pr_tag_img}
\end{figure*}

Tables~\ref{tab:clf_dr1} and \ref{tab:clf_dr1_tag} present additional results on the test set of the four multi-label classifiers (i.e. binary relevance with text, image and multi-modal information, and classifier chains with text data). 
The evaluation metrics 
are introduced in~\ref{appendix:metrics}.
We also report the precision-recall curves (on the test set) in Figure~\ref{fig:pr_tag_img} for each of the 11 tags produced by the multi-label classifier using multi-modal information.

\newpage
\begin{table}[htbp]
\centering
\caption{%
Classification performance evaluated with additional metrics.
\textbf{Text}: Binary relevance classifier using only text features; \textbf{Text (CC)}: Classifier chains using only text data; \textbf{Image}: Classifier using only images; \textbf{Text+Image}: Classifier using multi-modal information.
The best performance in terms of each metric (i.e. each row) is shown in \best{bold italic}.
}
\begin{tabular}{lrrrr}
\toprule
 & Text & Text (CC) & Image & Text+Image \\
\midrule
Balanced Accuracy  & 0.611 & 0.554 & 0.709 & \best{0.710} \\
macro-Precision    & 0.182 & 0.124 & 0.257 & \best{0.273} \\
macro-Recall       & 0.467 & \best{0.821} & 0.621 & 0.609 \\
macro-F1           & 0.246 & 0.207 & 0.350 & \best{0.367} \\
LRAP               & 0.496 & 0.476 & 0.689 & \best{0.693} \\
Coverage Error     & 4.003 & 4.248 & 2.730 & \best{2.679} \\
Label Ranking Loss & 0.256 & 0.278 & 0.130 & \best{0.127} \\
\bottomrule
\end{tabular}
\label{tab:clf_dr1}
\end{table}

\begin{table*}[htbp]
\centering
\caption{
Classification performance for text, image, and multi-modal information.
\textbf{Text}: Binary relevance classifier using only text features; \textbf{Text (CC)}: Classifier chains using only text data; \textbf{Image}: Classifier using only images; \textbf{Text+Image}: Classifier using multi-modal information.
The best performance in terms of each metric among 4 different multi-label classifiers is shown in \best{bold italic}.
}
\begin{tabular}{lrrrcrrrcrrrcrrr}
\toprule
\multirow{2}{*}{} & \multicolumn{3}{c}{Text} && \multicolumn{3}{c}{Text (CC)} && \multicolumn{3}{c}{Image} && \multicolumn{3}{c}{Text+Image} \\
\cmidrule{2-4} \cmidrule{6-8} \cmidrule{10-12} \cmidrule{14-16}
& Precision &  Recall &  F1  && Precision &  Recall &  F1  && Precision &  Recall &  F1 && Precision &  Recall &  F1 \\
\midrule
    artefact  & 0.043 & 0.312 & 0.076  && 0.014 & \best{0.500} & 0.027  && 0.062 & 0.312 & 0.104  && \best{0.066} & 0.312 & \best{0.109} \\
  asymmetric  & 0.162 & 0.435 & 0.236  && 0.116 & \best{0.983} & 0.208  && \best{0.165} & 0.565 & 0.256  && 0.164 & 0.593 & \best{0.257} \\
        bent  & 0.133 & 0.484 & 0.209  && 0.081 & \best{0.944} & 0.150  && 0.241 & 0.619 & 0.347  && \best{0.263} & 0.587 & \best{0.364} \\
     compact  & 0.288 & 0.436 & 0.347  && 0.238 & \best{0.947} & 0.380  && \best{0.548} & 0.792 & \best{0.648}  && 0.530 & 0.816 & 0.643 \\
      double  & 0.445 & 0.504 & 0.473  && 0.342 & \best{0.996} & 0.510  && 0.528 & 0.816 & \best{0.641}  && \best{0.531} & 0.796 & 0.637 \\
   hourglass  & 0.140 & 0.609 & 0.228  && 0.118 & \best{0.916} & 0.209  && 0.252 & 0.749 & \best{0.377}  && \best{0.254} & 0.704 & 0.373 \\
  noir\_ifrs  & 0.213 & 0.752 & 0.331  && 0.140 & \best{0.892} & 0.242  && 0.201 & 0.847 & 0.325  && \best{0.251} & 0.822 & \best{0.385} \\
    overedge  & 0.318 & 0.616 & 0.419  && 0.132 & \best{0.942} & 0.231  && 0.345 & 0.668 & 0.455  && \best{0.416} & 0.663 & \best{0.511} \\
   restarted  & 0.029 & 0.333 & 0.053  && 0.020 & \best{0.500} & 0.039  && 0.111 & 0.444 & \best{0.178}  && \best{0.120} & 0.333 & 0.176 \\
      triple  & 0.225 & 0.512 & 0.313  && 0.159 & \best{0.984} & 0.273  && 0.307 & 0.589 & 0.403  && \best{0.315} & 0.645 & \best{0.423} \\
     xshaped  & 0.011 & 0.143 & 0.021  && 0.006 & \best{0.429} & 0.011  && 0.068 & \best{0.429} & 0.118  && \best{0.097} & \best{0.429} & \best{0.158} \\
\bottomrule
\end{tabular}
\label{tab:clf_dr1_tag}
\end{table*}

\label{lastpage}
\end{document}